\documentclass{llncs}

\usepackage{times}
\usepackage{xspace}
\usepackage{amsmath}
\usepackage{amssymb}
\usepackage{url}
\usepackage{color}
\newcommand{\change}{\relax}

\newenvironment{zproof}[1]{\begin{proof}\begin{rm}#1}{\hfill$\Box$\end{rm}\end{proof}}

\newcommand{\infest}{\mathbin{|\kern -.47em\sim}}

\newcommand{\res}[1]{\mathord{|_{#1}}\xspace}

\begin{document}
\title{
Interpolation in Equilibrium Logic and Answer Set Programming: the Propositional Case}
\author{Dov Gabbay \inst{1}, David Pearce\inst{2}%
\thanks{Partially supported by CICyT projects TIC-2003-9001-C02 and TIN2006-15455-CO3.}
\and Agust\'\i n Valverde\inst{3}%
\thanks{Partially supported by CICyT project TIC-2003-9001-C01, TIN2006-15455-CO1 and Junta de Andalucia project TIC-115}}
\authorrunning{D. Pearce, A. Valverde}
\tocauthor{David Pearce (Universidad Politécnica de Madrid) and Agust\'\i n Valverde (Universidad de M\'alaga)}

\institute{King's College London\\
\email{dov.gabbay@kcl.ac.uk}
\and
AI Dept,
Universidad Polit\'ecnica de Madrid, Spain.\\
\email{david.pearce@upm.es}
\and
Dept of Applied Mathematics, Universidad de M\'alaga, Spain.\\
\email{a\_valverde@ctima.uma.es}
}

\maketitle

\begin{abstract}
Interpolation is an important property of classical and many non classical logics that has been shown to have interesting applications in computer science and AI. Here we study the Interpolation Property for the propositional version of the non-monotonic system of {\em equilibrium logic}, establishing weaker or stronger forms of interpolation depending on the precise interpretation of the inference relation. These results also yield a form of interpolation for ground logic programs under the answer sets semantics. For disjunctive logic programs we also study the property of {\em uniform} interpolation that is closely related to the concept of variable forgetting.
\end{abstract}

\section{Introduction}
The Interpolation Property plays an important role in logical systems, both classical and non-classical. Its importance in computer science has also become recognised  lately. 
The Interpolation Property has been applied in various areas of computer science, notably in software specification \cite{Bic01} and in the construction of formal ontologies \cite{KWZ08}. In both cases it is relevant to modularity issues, for instance in~\cite{KWZ08}
it plays a key role in the study of the modular decomposition of ontologies. However to date interpolation has received less attention in systems of nonmonotonic reasoning and logic programming, despite their importance in AI and computer science. In this paper we study the interpolation property for the system of nonmonotonic reasoning known as \emph{equilibrium logic}.%
\footnote{We are very grateful to Karl Schlechta for valuable discussions on the interpolation concept for non-monotonic logics.}
Since this in turn forms a logical foundation for stable model reasoning and Answer Set Programming (ASP), our results transfer easily to the sphere of ASP.

Let us start with some notation and terminology. Let us assume the syntax of \change{propositional} logic with formulas denoted by lower case Greek letters
% and predicates by lower case Latin letters. 
% We write $\alpha (\overline{p},\overline{q})$ to denote that predicates
% occurring in the formula $\alpha$ belong to $\overline{p}, \overline{q}$;
% similarly for other combinations of predicates. 
\change{and let $\vdash$ be a monotonic inference relation.
If  $\alpha \vdash \beta $, an \emph{interpolant for}}
$(\alpha, \beta)$ is a formula $\gamma$ such that
\begin{equation}
\label{eq:interpol}
\alpha  \vdash \gamma  \;\; \& \;\;  \gamma \vdash \beta
\end{equation}
where $\gamma$ contains only
% predicate and constant symbols 
\change{variables} that belong to both $\alpha$ and $\beta$. A logic $L$ with inference relation $\vdash_L$ is said to have the \emph{interpolation property} if an interpolant exists for every pair of formulas $(\alpha, \beta)$ such that $\alpha  \vdash_L \beta $. As is well-known, classical logic as well as many non-classical logics possess interpolation~\cite{GM2005}.

Suppose now we deal with a non-monotonic logical system with an inference relation $\infest$. To express the idea that a formula is an interpolant one approach is simply to replace $\vdash$ by $\infest$ in (\ref{eq:interpol}). 
%One problem is that, since $\infest$ is non-monotonic,  it is in general not transitive. 
We may call this form $(\infest, \infest)$-interpolation. 
%Instead, following the idea of \cite{GM2005},  we can modify condition (\ref{eq:interpol}) and proceed in a two-stage fashion. We make use of the fact that non-monotonic consequence can be defined in terms of minimal models in some monotonic logical system, say that the consequence relation $\infest$ is appropriately captured by means of minimal models in a logic $L$ with consequence relation $\models_L$. 
Although this is a legitimate form of interpolation, in practice we shall find it useful to study another approach suggested in  \cite{GM2005}.
Suppose that our non-monotonic inference can be defined in terms of minimal models in some monotonic logical system, say that the  relation $\infest$ is  captured by means of minimal models in a logic $L$ with inference relation $\vdash_L$\footnote{Assume by completeness that this coincides with $L$-consequence, $\models_L$.}.
Assume $\alpha  \infest \beta $;
then as an interpolant for $(\alpha, \beta)$ we look for a formula $\gamma$ such that
\begin{equation}
\label{eq:nonmon_interpol1}
\alpha \infest \gamma  \;\; \& \;\;  \gamma \vdash_L \beta 
\end{equation}
where
% all predicate and constant symbols
\change{all variables} of $\gamma$ occur in both $\alpha$ and $\beta$.
Since $\infest$ is to be defined via a subclass of minimal $L$-models, we already suppose that $\models_L \subseteq \infest$. Moreover we should require too that $L$ is a well-behaved sublogic in the sense that $L$-equivalent formulas have the same $\infest$-consequences and that formulas $L$-derivable from $\infest$-consequences are themselves $\infest$-consequences (so eg from (\ref{eq:nonmon_interpol1}) we can derive $\alpha \infest \beta$).
In non-monotonic reasoning these last two properties are known as left and right absorption, respectively~\cite{Ma94}.
Given these conditions, it follows at once from (\ref{eq:nonmon_interpol1}) that any formula in the language of $\gamma$ that is $L$-equivalent to $\gamma$ will also be an interpolant for $(\alpha, \beta)$. Likewise if $\gamma$ is an interpolant for $(\alpha, \beta)$ and   $ \beta  \vdash_L  \delta  $ then $\alpha \infest \delta$ and $\gamma$ is an interpolant for $(\alpha, \delta)$. When (\ref{eq:nonmon_interpol1}) holds we call $\gamma$ a $(\infest , \vdash_L)$-interpolant.

Now, to find a $(\infest , \vdash_L)$-interpolant for $(\alpha, \beta)$, we can proceed as follows. We look for an $L$-formula $\alpha'$ say, that precisely $L$-defines the minimal models of $\alpha$. Since \hbox{$\alpha \infest \beta$} it follows that \hbox{$\alpha' \models_L \beta $} and, assuming completeness,
\hbox{$\alpha' \vdash_L \beta $}.
Now, if $L$ has the interpolation property as defined earlier, we apply this theorem to obtain or infer the existence of an $L$-interpolant $\gamma$ in the sense of (\ref{eq:interpol}) for $(\alpha', \beta)$. Hence (\ref{eq:nonmon_interpol1}) follows.

\subsection{Scope of the paper}
In this paper we study interpolation for the propositional version of equilibrium logic, based on the non-classical, monotonic logic of {\em here-and-there}, $\mathbf{HT}$. We introduce two variants of equilibrium inference, denoted by $\infest_{cw}$ and $\infest_{ow}$ respectively. This notation reflects the idea that one form of inference is closer in spirit to closed-world reasoning while the other resembles more a type of open-world reasoning. While $(\infest, \vdash_{L})$-interpolation holds for $\infest_{ow}$ (setting $L = \mathbf{HT}$), only the weaker form of
$(\infest, \infest)$-interpolation holds for $\infest_{cw}$. However in both cases we apply the general method described above, using definable classes of minimal $\mathbf{HT}$-models.

The restriction to propositional equilibrium logic is quite sufficient for considering interpolation in ASP, for the case of finite, ground logic programs (of any syntactic kind). In the case of ASP, the most natural associated form of inference would seem to be $\infest_{cw}$, satisfying $(\infest, \infest)$-interpolation. As a final topic we consider the extent to which a stronger form of {\em uniform} interpolation holds for disjunctive programs under a restricted query language. Here we make use of recent results by Eiter and Wang~\cite{EW08} on variable forgetting in ASP.

For reasons of space we do not consider here the full, first-order version of equilibrium logic that can serve as a foundation for non-ground answer set programs. This is done in the full, extended version of the paper that is currently in preparation.

\section{Logical Preliminaries }

%\subsection{Logics and interpolation}

We work with standard propositional languages, $\mathcal{L}$, $\mathcal{L'}$, etc
%, where the latter may in the general case contain constant and function symbols.
% Propositional languages are
based on a sets $V$, $V'$, of propositional variables.\footnote{Many logic texts work with a fixed, countable set of propositional variables. However here we find it useful to distinguish different languages $\mathcal{L}$, $\mathcal{L'}$, and variable sets $V,V'$, etc. For one thing, these languages may arise by grounding or instantiating finite, first-order theories, as occurs in ASP, hence it may be important to distinguish the different languages that may result from this process. Secondly, in a non-monotonic context, as we shall see, different kinds of inference relations may arise according to the way in which language extensions are handled. Thirdly, our definitions will be easily extended to the first-order case.}
Formulas are built-up in the usual way using the logical constants $\land$, $\lor$, $\to$, $\neg$, standing respectively for conjunction, disjunction, implication and  negation.
If $\varphi$ is a propositional formula, we denote by $V(\varphi)$ the set of propositional variables appearing in $\varphi$.
%
%A \emph{first-order language} $\mathcal{L}=\langle C, F , P \rangle$ consists of a set of constants $C$, function symbols $F$ and predicate symbols $P$;
%each function symbol $f$ in $F$ and predicate symbol $p\in P$ has an assigned arity.
%Moreover, we assume a fixed countably infinite set of variables, the symbols, `$\to$', `$\lor$', `$\land$', `$\neg$', `$\exists$', `$\forall$' and auxiliary parentheses `$($',`$)$'.
%\emph{Atoms}, \emph{terms}  and \emph{formulas} are constructed as usual;
%\emph{closed} formulas, or \emph{sentences}, are those where each variable is bound by some quantifier. If $\varphi$ is a (first-order) formula, $\mathcal{L} (\varphi)$ denotes the language associated with $\varphi$, ie the set of constants, function and predicate symbols occuring in it.
%
%We make use of the following notation and terminology. Boldface $\mathbf{x}$ stands for a tuple of variables,
%$\mathbf{x}=(x_1,\dots,x_n)$, while 
%$\varphi(\mathbf{x})=\varphi(x_1,\dots,x_n)$ is a formula whose free variables are $x_1$,\dots, $x_n$, and $\forall\mathbf{x}=\forall x_1\dots\forall x_n$.
%If $t_i$ are terms, then $\mathbf{t}=(t_1,\dots,t_n)$ denotes a \emph{vector} of terms.
%A \emph{theory} $\Pi$ is a set of sentences. Variable-free terms, atoms, formulas, or theories are also called \emph{ground}.

As usual the symbols $\vdash$ and $\models$, possibly with subscripts, are used to denote logical inference and consequence relations, respectively.
A logic $L$ is said to be \emph{monotonic} if its inference relation $\vdash_L$ satisfies the monotonicity property:
\begin{equation}
\Pi \vdash_L \varphi \;\; \& \;\; \Pi \subseteq \Pi' \; \to \; \Pi' \vdash_L \varphi
\end{equation}

To distinguish non-monotonic from monotonic inference relations, we use $\infest$ to symbolise the former. In most cases a non-monotonic logic can be understood in terms of an inference relation that extends a suitable monotonic logic.
When this extension is well-behaved we say that the monotonic logic forms a \emph{deductive base} for it.
This can be made precise as follows.
\begin{definition}\label{ded-base}
Let $\infest$ be any nonmonotonic inference relation. We say that a logic ${L}$ with monotonic inference relation $\vdash_{L}$ is a \emph{deductive base} for $\infest$ iff (i) $\vdash_{L} \subseteq \infest$; (ii)  If $\Pi_1 \equiv_{L} \Pi_2$ then $\Pi_1 \approx \Pi_2 $; (iii) If $\Pi \infest \varphi$ and $\varphi \vdash_{L} \psi$, then $\Pi \infest \psi$.
\end{definition}
Here $\equiv_{L}$ denotes ordinary logical equivalence in $L$, while $\approx$ denotes non-monotonic equivalence, ie $\Pi_1 \approx \Pi_2 $ means that $\Pi_1$ and $\Pi_2$ have the same non-monotonic consequences.
Furthermore, we say that a deductive base is \emph{strong} if it satisfies the additional condition:
\[
\Pi_1 \not \equiv_{L} \Pi_2\quad \to \quad
\mbox{ there exists }
\Gamma \mbox{ such that }  \Pi_1 \cup \Gamma  \not \approx \Pi_2 \cup \Gamma.
\]
In terms of nonmonotonic consequence operations, (ii) and (iii) correspond to conditions known as left absorption and right absorption respectively, see \cite{Ma94}.\footnote{In the terminology of \cite{Di94} we therefore require of $\vdash_{L}, \infest$ that they form a \emph{fully absorbing inferential frame}.} 

We now turn to the interpolation property. 
\begin{definition}\label{def:interpolation}
A logic $L$ with inference relation $\vdash_L$ is said to have the \emph{interpolation property} if whenever
\[
\vdash_L \varphi \to \psi
\]
there exists a sentence $\xi$ (the \emph{interpolant}) such that
$V(\xi ) \subseteq V(\varphi ) \cap V( \psi )$ and
\[
\vdash_L \varphi \to \xi \quad\text{and}\quad\vdash_L \xi \to \psi
\]
% where all predicate, function and constant symbols of $\xi$ are contained
% in both $\varphi$ and $\psi$, ie $\mathcal{L(\xi)} \subseteq
% \mathcal{L(\varphi)} \cap \mathcal{L(\psi)}$. In the case of propositional logic,
% the requirement is that
\end{definition}

As explained in the introduction, for non-monotonic logics we can consider two forms of interpolation, one weaker one stronger. The stronger form makes use of an underlying monotonic logic.
\begin{definition}
Suppose that  $\alpha \infest \beta $. A $(\infest, \vdash_L )$ \emph{interpolant for} $(\alpha, \beta)$ is a formula $\gamma$ such that $V(\gamma) \subseteq V(\alpha) \cap V(\beta)$ and
\begin{equation}\label{eq:interpol1}
\alpha  \infest \gamma \quad\text{and}\quad\gamma \vdash_L \beta
\end{equation}
where $L$ is a deductive base for $\infest$.
% $\gamma$ contains only predicate, function and constant symbols
% that belong to both $\alpha$ and $\beta$.
% $V(\gamma) \subseteq V(\alpha) \cap V(\beta)$.
A non-monotonic logic  with inference relation $\infest$ is said to have the $(\infest,\vdash)$ \emph{interpolation property} if for a suitable deductive base logic $L$ a $(\infest, \vdash_L )$ interpolant exists for every pair of formulas $(\alpha, \beta)$ such that $\alpha  \infest \beta $.
\end{definition}
The requirement that $L$ form a deductive base ensures that some desirable properties of interpolation are met. 
\begin{proposition}
\label{prop:interprops}
Let $\gamma$ be a $(\infest, \vdash_L )$ \emph{interpolant for} $(\alpha, \beta)$.
\begin{enumerate}
\item
For any $\psi$ such that $\psi \equiv_L \gamma$, $\psi$ is a $(\infest, \vdash_L )$ \emph{interpolant for} $(\alpha, \beta)$. 
\item
For any $\alpha'$ such that $\alpha \equiv_L \alpha'$, and any $\beta'$ such that $\beta \vdash_L \beta'$, $\gamma$ is a $(\infest, \vdash_L )$ \emph{interpolant for} $(\alpha', \beta')$.
\end{enumerate}
\end{proposition}
The property of deductive base also guarantees that the $(\infest, \vdash_L )$ relation is transitive in the sense that if (\ref{eq:interpol1}) holds for any $\alpha, \beta , \gamma$, then also $\alpha \infest \beta$.
This last property will not necessarily hold for the second, weaker form of interpolation that we call $( \infest , \infest )$ interpolation.
\begin{definition}
\label{def:interpol2}
Suppose that  $\alpha \infest \beta $. A $(\infest, \infest )$ \emph{interpolant for} $(\alpha, \beta)$ is a formula $\gamma$ such that $V(\gamma) \subseteq V(\alpha) \cap V(\beta)$ and
\begin{equation}
\label{eq:interpol2}
\alpha  \infest \gamma \quad\text{and}\quad \gamma \infest \beta
\end{equation}
% where $\gamma$ contains only predicate, function and constant symbols that
% belong to both $\alpha$ and $\beta$.
% In the case of propositional logic, the requirement is that
\end{definition}
Analogous to the previous case, we say that a non-monotonic logic  with inference relation $\infest$ has the $(\infest, \infest )$ \emph{interpolation property} if a $(\infest, \infest )$ interpolant exists for every pair of formulas $(\alpha, \beta)$ such that $\alpha  \infest \beta $. Notice that $(\infest, \vdash_L )$ is the stronger form of interpolation because if a logic has $(\infest, \vdash_L )$ interpolation it must also have $(\infest, \infest )$ interpolation, again as a consequence of the deductive base requirement (first clause). 

Evidently the properties expressed in Proposition \ref{prop:interprops} are not directly applicable to the second form of interpolation that does not refer to any underlying base logic. Nevertheless an important feature of the interpolation properties we shall establish below is that we can formulate and prove analogous properties even for $(\infest, \infest )$ interpolation.

We can also consider restricted variants of interpolation when the property holds for certain types of formulas, in other words, when there is an interpolant for $(\alpha, \beta)$
% given $\alpha  \infest \beta $
\emph{whenever $\alpha$ and $\beta$ belong to specific syntactic classes}. 
%In such cases we can refer to \emph{interpolable} formulas. 
Later on we shall consider both kinds of restrictions, where
$\alpha$ belongs to a specific class or alternatively when $\beta$ does. 

\subsection{Review of the Logic of Here-and-There}

Equilibrium logic is based on the nonclassical logic of here-and-there,  which we denote by $\mathbf{HT}$ in the propositional case.
The axioms and rules of inference for $\mathbf{HT}$ are those of intuitionistic logic together with the axiom schema 
\[
\alpha \lor (\neg\beta \lor (\alpha\to \beta)).
%(\neg \alpha \to  \beta ) \to
%(((\beta \to \alpha ) \to \beta ) \to \beta)
\]
The model theory of $\mathbf{HT}$ is based on the usual Kripke semantics for intuitionistic logic~\cite{Da83}, but it is complete for Kripke frames with just two worlds $h$ (here) and $t$ (there) such that $h\leq t$. We can therefore represent a Kripke model for $\mathbf{HT}$ as a triple $\langle \{ h , t \} , \leq , v \rangle$, where $v$ is a (truth) valuation. Alternatively, we can speak of an $\mathbf{HT}$-{\em interpretation} as an ordered pair $\mathcal{M} = \langle H, T \rangle$ of sets of atoms such that $H \subseteq T$;
the elements of $H$ are the atoms true \emph{here} and the elements of $T$ are the atoms true \emph{there}. The truth of a formula $\varphi$ in a world $\mathcal{M},w\models \varphi$ is defined recursively via the usual rules for conjunction, disjunction, implication and negation in intuitionistic logic.
%\begin{itemize}
%\item
%$\mathcal{I},w\models \varphi \land \psi$ iff $\mathcal{I},w\models
%\varphi$ and $\mathcal{I},w\models \psi$.
%\item
%$\mathcal{I},w\models \varphi \lor \psi$ iff $\mathcal{I},w\models
%\varphi$ or $\mathcal{I},w\models \psi$.
%\item
%$\mathcal{I},t\models \varphi \to \psi$ iff $\mathcal{I},t
%\not\models \varphi$ or $\mathcal{I},t \models \psi$.
%\item
%$\mathcal{I},h\models \varphi \to \psi$ iff $\mathcal{I},t\models
%\varphi \to \psi$ and $\mathcal{I},h \not\models \varphi$ or
%$\mathcal{I},h \models \psi$.
%\item
%$\mathcal{I},w\models \neg \varphi$ iff $\mathcal{I},t \not\models
%\varphi$.
%\item
%$\mathcal{I},t \models \forall x \varphi(x)$ iff $\mathcal{I},t
%\models\varphi(d)$ for all  $d\in D$.
%\item
%$\mathcal{I},h \models \forall x \varphi(x)$ iff $\mathcal{I},t
%\models \forall x \varphi(x)$ and  $\mathcal{I}, h \models
%\varphi(d)$ for all  $d\in D$.
%\item
%$\mathcal{I},w\models \exists x \varphi(x)$ iff
%$\mathcal{I},w\models\varphi(d)$ for some $d\in D$.
%\end{itemize}
A formula $\varphi$ is true in $\mathcal{M} = \langle H, T \rangle$ in symbols $\mathcal{M}\models\varphi$, if it is true at each world in $\mathcal{M}$;
in this case we say that $\mathcal{M}$ is an $\mathbf{HT}$-model of $\varphi$.
A formula $\varphi$ is said to be \emph{valid} in $\mathbf{HT}$, in symbols $\models\varphi$, if it is true in all $\mathbf{HT}$-interpretations.
Logical consequence for $\mathbf{HT}$ is understood as follows:
$\varphi$ is said to be an $\mathbf{HT}$ consequence of a theory $\Pi$, written $\Pi \models \varphi$, iff for all models $\mathcal{M}$ and any world $w\in\mathcal{M}$, $\mathcal{M},w\models\Pi$ implies $\mathcal{M},w\models\varphi$.
Equivalently this can be expressed by saying that $\varphi$ is true in all $\mathbf{HT}$-models of $\Pi$.

More exactly, we might to write $\mathbf{HT}(\mathcal{L})$ to refer to the language considered in the logic.
However, as we see below, the logic is in fact independent from the language.

Let $\mathcal{L}$ be a proper sublanguage of $\mathcal{L'}$, ie $\mathcal{L}\subset\mathcal{L'}$; for any $\mathbf{HT}(\mathcal{L}')$-interpretation $\mathcal{M}= \langle H,T \rangle$ we denote by
$\mathcal{M}\res{\mathcal{L}}$ the $\mathbf{HT}(\mathcal{L})$-interpretation formed by omitting the interpretation of all atoms in $\mathcal{L'}\smallsetminus\mathcal{L}$ and we call this the \emph{reduct} of $\mathcal{M}$ to $\mathcal{L}$.
\begin{proposition}
Suppose that $\mathcal{L}'\supset \mathcal{L}$, $\Pi$ is a theory in $\mathcal{L}$
and $\mathcal{M}$ is an $\mathbf{HT}(\mathcal{L}')$-model of $\Pi$. Then $\mathcal{M}\res{\mathcal{L}}$ is a $\mathbf{HT}(\mathcal{L})$-model of $\Pi$.
\end{proposition}
\begin{proposition}
Suppose that $\mathcal{L}'\supset \mathcal{L}$ and  $\varphi\in\mathcal{L}$.
Then $\varphi$ is valid (resp. satisfiable) in $\mathbf{HT}(\mathcal{L})$ if and only if is valid (resp. satisfiable) in $\mathbf{HT}({\mathcal{L}'})$.
\end{proposition}

A key property of here-and-there is that, as Maksimova~\cite{Ma77} showed, it is one of just seven super-intuitionistic logics with Interpolation.

\begin{proposition}[\cite{Ma77}]
The logic $\mathbf{HT}$ possesses the Interpolation Property.
\end{proposition}

\subsection{Equilibrium Logic}

Equilibrium logic is based on certain kinds of minimal models in $\mathbf{HT}$. 

\begin{definition}
Among here-and-there interpretations we define the order $\unlhd$ as follows:
$\langle H,T\rangle\unlhd\langle H',T'\rangle$ if $T=T'$ and $H\subseteq H'$.
If
% the subset relation holds strictly, 
$H\subset H'$ we write
% `$\lhd$'.
$\langle H,T\rangle\lhd\langle H',T'\rangle$
\end{definition}

\begin{definition}[Equilibrium model]
Let $\Pi$ be a theory and $\mathcal{M}=\langle H,T\rangle$ a model of~$\Pi$.
\begin{enumerate}
\item
$\mathcal{M}$ is said to be \emph{total} if $H=T$.
\item
$\mathcal{M}$ is said to be an \emph{equilibrium} model of $\Pi$ if it is minimal under $\unlhd$ among models of $\Pi$, and it is total.
% It is denoted by $\mathcal{I}\infest \Pi$.
\end{enumerate}
\end{definition}
In other words, equilibrium models are total models for which there is no `smaller' non-total model. Evidently a total $\mathbf{HT}$-model of a theory $\Pi$ can be equivalently regarded as a classical model of $\Pi$;
and in what follows we make tacit use of this equivalence. A theory is said to be {\em consistent} if it has an $\mathbf{HT}$-model and {\em coherent} if it has an equilibrium model.

We define a preliminary notion of equilibrium entailment as follows. It essentially agrees with standard versions of equilibrium logic, as eg in \cite{Pe06}.
\begin{definition}\label{def:eq-entailment}
The relation $\infest$, called \emph{equilibrium entailment}, is defined as follows.
Let $\Pi$ be a set of formulas.
\begin{enumerate}
\item
If $\Pi$ is non-empty and has equilibrium
models, then $\Pi \infest \varphi$ if
every equilibrium model of $\Pi$ is a model of $\varphi$ in $\mathbf{HT}$.
\item
If either $\Pi$ is empty or  has no equilibrium models, then $\Pi \infest \varphi$ if $\Pi \vdash \varphi$.
\end{enumerate}
\end{definition}
A few words may help to explain the concept of equilibrium entailment. First, we define the basic notion of entailment as truth in every intended (equilibrium) model.
In nonmonotonic reasoning this is a common approach and sometimes called a \emph{skeptical} or \emph{cautious} notion of entailment or inference;
its counterpart \emph{brave} reasoning being defined via truth in some intended model.
Since equilibrium logic is intended to provide a logical foundation for the answer set semantics of logic programs, the cautious variant of entailment is the natural one to choose:
the standard consequence relation associated with answer sets is given by truth in all answer sets of a program.
Note however that in ASP as a programming paradigm each answer set may correspond to a particular solution of the problem being modelled and is therefore of interest in its own right.

Secondly, it is useful to have a nonmonotonic consequence or entailment relation that is non-trivially defined for all consistent theories.
As is easily seen, however, not all such theories possess equilibrium models.
For such cases it is natural to use monotonic consequence as the entailment relation.
In particular, $\mathbf{HT}$ is a maximal logic with the property that logically equivalent theories have the same equilibrium models.
Evidently situation $2$ in previous definition also handles correctly the cases that $\Pi$ is empty or inconsistent. 

Despite these qualifications, there remains an ambiguity in the concept of equilibrium entailment that we now need to settle.
Suppose that $\mathcal{L}'\supset \mathcal{L}$, $\Pi$ is a theory in $\mathcal{L}$
and $\varphi$ is a sentence in $\mathcal{L}'$.
How should we understand the expression `$\Pi \infest \varphi$'? 

Evidently, if we fix a language in advance, say as the language $\mathcal{L}'$, then we can simply consider the equilibrium models of $\Pi$ in $\mathcal{L}'$.
But if $\Pi$ represents a knowledge base or a logic program, for instance, we may also take the view that $V(\Pi)$ is the appropriate language to work with.
In that case, the query $\varphi$ is as such not fully interpreted as it contains some variables not in $V(\Pi)$.

For any language $\mathcal{L}$ and $\mathcal{L}$-theory $\Pi$, let $E_{\mathcal{L}}(\Pi)$ be the collection of all equilibrium models of $\Pi$ in $\mathbf{HT}(\mathcal{L})$.
Now consider the following two variants of entailment. 
\newcommand{\extend}{\mathord{\upharpoonright}}
\begin{definition}[Equilibrium entailment]
\label{def:cw-ow}
Assume $\Pi$ is non-empty and has equilibrium models, then:
\begin{enumerate}
\item
Let us say that $\Pi \infest_{cw} \varphi$ if and only if $\mathcal{M} \models \varphi$ for each $\mathcal{M} \in E_{\mathcal{L}'}(\Pi)$, where $\mathcal{L}'$ is the language over $V(\Pi\cup\{\varphi\})$.
\item
Let us say that $\Pi \infest_{ow} \varphi$ if and only if $\mathcal{M} \models \varphi$ for each $\mathcal{M} \in E_{\mathcal{L}}(\Pi) \extend^{\mathcal{L}'}$, where $\mathcal{L}'$ is the language over $V(\varphi)$ and $E_{\mathcal{L}}(\Pi) \extend^{\mathcal{L'}}$ denotes the collection of all expansions of elements of $E_{\mathcal{L}} (\Pi)$ to models in $\mathcal{L} \cup \mathcal{L'}$, ie where the vocabulary of $\mathcal{L'} \smallsetminus \mathcal{L}$ is interpreted arbitrarily.
\end{enumerate}
\end{definition}

Obviously, if either $\Pi$ is empty or  has no equilibrium models, then $\Pi \infest_{cw} \varphi$ iff $\Pi \infest_{ow} \varphi$ iff $\Pi \vdash \varphi$.

A simple example will illustrate the difference between $\infest_{cw}$ and $\infest_{ow}$. Let $\psi$ be an $\mathcal{L}$-formula and let $q$ be a variable not in $\mathcal{L}$ and let $\mathcal{L}'$ be the language $\mathcal{L} \cup \{ q \}$.
By the first method we have $\psi \infest_{cw} \psi \land (q \lor \neg q)$.
In fact we have the stronger entailment $\psi \infest_{cw} \psi \land \neg q$.
The reason is that when we form the equilibrium models of $\psi$ in $\mathcal{L}'$, $q$ will be false in each as an effect of taking minimal models.
On the other hand, if we expand equilibrium models of $\psi$ in $\mathbf{HT}(\mathcal{L})$ to $\mathbf{HT}(\mathcal{L}')$, the new variable $q$ receives an arbitrary interpretation in
$\mathbf{HT}(\mathcal{L}')$.
Since this logic is 3-valued we do not obtain $ \Pi \infest_{ow} q \lor \neg q$.

For standard, monotonic logics, there is no difference between these two forms of entailment. If in Definition \ref{def:cw-ow} we replace everywhere equilibrium model by simply model (in $\mathbf{HT}$), variants (i) and (ii) give the same result.

In the context of logic programming and deductive databases the more orthodox view is that reasoning is based on a \emph{closed world assumption} (CWA). Accordingly a ground atomic query like $q(a)$?, where the predicate $q$ or the name $a$ do not belong to the language of the program or database would simply be assigned the value \emph{false}. This is also the case with the first kind of equilibrium entailment and we use the label $\infest_{cw}$ since this variant appears closer to a closed world form of reasoning. On the other hand, there may be legitimate cases where we do not want to apply the CWA and where unknown values should be assigned to  an atom that is not expressed in the theory language. Then the second form of entailment, $\infest_{ow}$, nearer to open world reasoning, may then be more appropriate. For present purposes, however, the suffices `$cw$' and `$ow$' should be thought of merely as mnemonic labels.
A more thorough analysis of closed world versus open world reasoning in this context would probably lead us to consider assumptions such as \emph{unique names assumption} or \emph{standard names assumption} and is outside the scope of this paper.

\section{Interpolation in Propositional Equilibrium Logic}

In this section we deal with interpolation in propositional equilibrium logic. It is clear that by its semantic construction propositional equilibrium logic has $\mathbf{HT}$ as a deductive base. This base is actually maximal.
\begin{proposition}\label{ded_base}
$\mathbf{HT}$ is a strong and maximal deductive base for (propositional) equilibrium entailment.
\end{proposition}
The first property is precisely the strong equivalence theorem of \cite{LPV01}. Maximality follows from the fact that any logic strictly stronger than $\mathbf{HT}$ would have to contain classical logic which is easily seen not to be a deductive base, eg violating condition (ii) of Definition \ref{ded-base}.
We have:

\begin{lemma}
\label{lemma:definable}
Let $\alpha$ be a coherent $\mathbf{HT}$-formula and $E(\alpha)$ its set of equilibrium models. Then there is  formula $\alpha'$ of $\mathbf{HT}$ in $V(\alpha)$ that defines $E(\alpha)$ in the sense that $\mathcal{M} \in E(\alpha)$ iff
$\mathcal{M} \models \alpha'$.
\end{lemma}
\begin{zproof}
Suppose that $\alpha$ is coherent. and let 
\[ \mathcal{M}_1 = \langle T_1 , T_1 \rangle,  \mathcal{M}_2 = \langle T_2 , T_2 \rangle , \ldots , \mathcal{M}_n = \langle T_n , T_n \rangle \] 
be an enumeration of its equilibrium models. We show how to define $E(\alpha)$. Suppose each $T_i$, has $k_i$ elements and denote them by $A^{i}_{1}, \ldots , A^{i}_{j} , \ldots , A^{i}_{ k_{i}}$. Let $\overline{T_{i}}$ be the complement of $T_i$; then we can list its members as  
$A^{i}_{k_{1} + 1}, \ldots A^{i}_{l} \ldots , A^{i}_{ 
\mathbin{|} V (\alpha) \mathbin{| }}$. Set
\begin{equation}
\delta^{i} = \bigwedge_{j = 1, \ldots , k_i } A^{i}_{j} \land \neg ( \bigvee_{l = k_{i+1} , \ldots ,  \mathbin{|} V (\alpha) \mathbin{| }} A^{i}_{l} )
\end{equation}
Now set
\begin{equation}
\alpha' = \bigvee_{i = 1 , \ldots , n} \delta^{i}
\end{equation}
We claim that $\mathcal{M} \models \alpha'$ if and only if $\mathcal{M} = \mathcal{M}_i$ for some $i = 1, \ldots , n$, ie the models of $\alpha'$ are precisely $\mathcal{M}_1 , \ldots , \mathcal{M}_n$. To verify this claim, note that each $\mathcal{M}_i \models \delta^{i}$ and so $\mathcal{M}_i \models \alpha'$. Conversely, suppose that $\mathcal{M} \models \alpha'$. From the semantics of $\mathbf{HT}$ it is clear that $\mathcal{M} \models \varphi \lor \psi$ iff $\mathcal{M} \models \varphi$ or $\mathcal{M} \models  \psi$, so in particular $\mathcal{M} \models \alpha'$ implies $\mathcal{M} \models \delta^{i}$ for some $i = 1 , \ldots , n $. However, each $\delta^{i}$ defines a complete theory whose models are total. It follows that if $\mathcal{M} \models \delta^{i}$, then $\mathcal{M} = \mathcal{M}_i$. This establishes the claim.
\end{zproof}

Although we shall now demonstrate interpolation in the ($\infest, \infest$) form for the relation $\infest_{cw}$, we actually establish a stronger result. One consequence of this is that if we are concerned  with $\infest_{ow}$ entailment then the $(\infest, \vdash)$ form of interpolation actually holds.
%%%%%%%changed order%%%%%
\begin{proposition}[$\infest, \infest$-Interpolation]
\label{prop:interpolCW}
Let $\alpha, \beta$ be formulas and set $V = V(\alpha) \cup V(\beta)$ and $V' = V(\beta ) \smallsetminus V(\alpha) $ and suppose that $B_1 , \ldots B_n$ is an enumeration of $V'$.
If $\alpha \infest_{cw} \beta$, there is a formula $\gamma$ such that $V(\gamma) \subseteq V(\alpha) \cap V(\beta) $,  $\alpha \infest \gamma$, and $\gamma \land \neg B_1 \land \ldots \land \neg B_n \models \beta$. Hence in particular $\gamma \infest_{cw} \beta$.
\end{proposition}
\begin{zproof}
 Let $\alpha, \beta$ and $V,V'$ be as in the statement of the proposition, and suppose that $\alpha \infest_{cw} \beta$. Then $\beta$ holds in all equilibrium models of $\alpha$ in the language $V$. Case (i): suppose that $\alpha$ is coherent and form its set of equilibrium models, $E_{V}(\alpha)$.  
% As in the proof of Proposition \ref{prop:interpol} consider set $E(\alpha$
% of  equilibrium models of $\alpha$, considered as models in the language $V$. 
By the equilibrium construction it is easy to see that in each  model $\mathcal{M} \in E_V (\alpha)$ each atom $B_i$ is false, for $i = 1, n$. Construct the formulas $\delta_i$ and the formula $\alpha'$ exactly as in the proof of Lemma \ref{lemma:definable}. Now consider the formula $(\neg B_1 \land \ldots \land \neg B_n) \land \alpha'$. Clearly this formula defines the set of equilibrium models of $\alpha$ in $\mathbf{HT} (V)$. Consequently, 
$(\neg B_1 \land \ldots \land \neg B_n) \land \alpha' \models \beta$ and so $\alpha' \vdash (\neg B_1 \land \ldots \land \neg B_n) \to \beta$. We can now apply the interpolation theorem for $\mathbf{HT}$ to infer that there is a formula $\gamma$ such that $\alpha' \vdash \gamma$ and $\gamma \vdash (\neg B_1 \land \ldots \land \neg B_n) \to \beta$, where $V(\gamma) \subseteq V(\alpha' ) \cap V(\beta)$ and hence $V(\gamma) \subseteq V(\alpha ) \cap V(\beta)$. Since $\mathbf{HT}$ is a deductive base, we conclude that
\[
  \alpha \infest \gamma \;\; \& \;\; \gamma \land \neg B_1 \land \ldots \land \neg B_n \vdash \beta .
\]
Now, since $V(\gamma) \subseteq V(\alpha ) \cap V(\beta)$, $B_i \not \in V(\gamma)$ for $i = 1,\dots, n$. It follows that in $\mathbf{HT} (V(\beta))$, each $B_i$ is false in every equilibrium model of $\gamma$. So each such model $\mathcal{M}$ satisfies $( \neg B_1 \land \ldots \land \neg B_n)$.\footnote{Notice that in this case adding to $\gamma$ the sentence $( \neg B_1 \land \ldots \land \neg B_n)$ does not change its set of equilibrium models.} Since each also satisfies $\beta$, we have $\gamma \infest_{cw} \beta$.

Case (ii). If $\alpha$ has no equilibrium models then the hypothesis is that $\alpha \vdash \beta$. In that case we simply choose an interpolant $\gamma$ for $(\alpha, \beta)$.
\end{zproof}
%%%%%%%%%%%%%%%%%%%%%%%%%

\begin{corollary}[($\infest, \vdash$)-Interpolation]
\label{prop:interpol}
\hspace{.2em} Let $\alpha, \beta$ be formulas such that $\alpha \infest_{cw} \beta$ and
\hbox{$V(\beta) \subseteq V(\alpha)$}. There is a formula $\gamma$ such that $V(\gamma) \subseteq V(\alpha) \cap V(\beta) $ and  $\alpha \infest_{cw} \gamma$ and $\gamma \vdash \beta$.
\end{corollary}
\begin{zproof}
Immediate from Proposition \ref{prop:interpolCW} by the fact that $V(\beta) \smallsetminus V(\alpha) = \emptyset$.
\end{zproof}

\begin{proposition}[($\infest, \vdash$)-Interpolation]
\label{prop:interpolOW}
Let $\alpha, \beta$ be formulas and set $V = V(\alpha) \cup V(\beta)$ and $V' = V(\beta ) \smallsetminus V(\alpha) $.  If $\alpha \infest_{ow} \beta$, there is a formula $\gamma$ such that $V(\gamma) \subseteq V(\alpha) \cap V(\beta) $,  $\alpha \infest_{ow} \gamma$, and $\gamma  \vdash \beta$. 
\end{proposition}
\begin{zproof}
 Let $\alpha, \beta$ and $V,V'$ be as in the statement of the proposition and suppose that $\alpha \infest_{ow} \beta$. Then $\beta$ holds in all expansions of elements of $E_{V(\alpha)} (\alpha)$ to the language $V$. Case (i): suppose that $\alpha$ is coherent and consider $E_{V(\alpha) }(\alpha)$.  
% As in the proof of Proposition \ref{prop:interpol} consider set
% $E(\alpha$ of  equilibrium models of $\alpha$, considered as models in the language $V$. 
Again construct the formulas $\delta_i$ and the formula $\alpha'$ exactly as in the proof of Lemma \ref{lemma:definable}. Now consider $\alpha'$ which defines the set $E_{V(\alpha) }(\alpha)$. Then $\beta$ holds in all expansions of models of $\alpha'$ to $V$. Hence $\alpha' \models \beta$ and therefore $\alpha' \vdash \beta$
 We can now apply the interpolation theorem for $\mathbf{HT}$ to infer that there is a formula $\gamma$ such that $\alpha' \vdash \gamma$ and $\gamma \vdash \beta$, where $V(\gamma) \subseteq V(\alpha' ) \cap V(\beta)$ and hence $V(\gamma) \subseteq V(\alpha ) \cap V(\beta)$. Since $\alpha \infest_{ow} \alpha'$ and $\mathbf{HT}$ is a deductive base we conclude that
\[
  \alpha \infest_{ow} \gamma \;\; \& \;\; \gamma  \vdash \beta .
\]
Case (ii). If $\alpha$ has no equilibrium models, choose $\gamma$ as an interpolant for $(\alpha, \beta)$.
\end{zproof}

\section{Interpolation in Answer Set Semantics}

Answer set programming (ASP) has become an established form of declarative, logic-based programming and its basic ideas are now well-known. For a textbook treatment the reader is referred to \cite{bara-2002}. As is also well-known, the origins of ASP lie in the stable model and answer set semantics for logic programs introduced
% Stable models for normal logic programs, whose rules have atomic heads
% and may contain default negation in their bodies, were defined 
in \cite{GL88}. This semantics made use of a fixpoint condition involving a certain `reduct'  operator. Subsequent extensions of the concept to cover more general kinds of rules also relied on a reduct operator of similar sort~\cite{LTT99,Fe05}. For the original definitions, the reader is referred to the various papers cited.

The correspondence between answer set semantics and equilibrium logic is also well-established and has been discussed in many publications, beginning with \cite{Pe97} which first showed how the answer sets of disjunctive programs can be regarded as equilibrium models. 
%For subsequent developments of this idea see \cite{LPV01,LPV07,fer07a,PV05,PV06,PV08}. 
For our purposes it will suffice to recall just the main features of the correspondence with equilibrium logic.

We recall the notion of ground, disjunctive logic program (without strong negation) whose answer sets are simply collections of atoms.
These programs consist of sets of ground rules of the form
\begin{equation}
\label{disj_rule}  K_1 \lor \ldots \lor K_k \leftarrow L_1, \ldots  L_m, not L_{m+1}, \ldots , not L_n 
\end{equation}
where the $L_i$ and $K_j$ are atoms. The `translation' from the syntax of programs to $\mathbf{HT}$ propositional formulas is the trivial one, viz. (\ref{disj_rule}) corresponds to the $\mathbf{HT}$ sentence
\begin{equation}
\label{disj_formula}
 L_1 \land \ldots  \land L_m \land \neg L_{m+1} \land \ldots \land \neg L_n \to K_1 \lor \ldots \lor K_k 
 \end{equation}
Under this translation the correspondence between the answer sets and the equilibrium models of ground disjunctive programs is also the direct one:
\begin{proposition}\label{AS}
Let $\Pi$ be a disjunctive logic program. Then $\langle T,T \rangle$ is an equilibrium model of $\Pi$ if and only if $T$ is an answer set of $\Pi$.
\end{proposition}
This was first shown  in \cite{Pe97}, but the basic equivalence was later shown to hold for more general classes of programs  in \cite{LPV01}. Indeed it can also be extended to embrace the very general definition of answer set for propositional theories, given by Ferraris~\cite{Fe05}.

In ASP the main emphasis is on finding answer sets and this is what most answer set solvers compute. Less attention is placed on implementing a non-monotonic inference relation.\footnote{For example the main solvers such as {\tt smodels}, DLV or CLASP do not implement a query answering mechanism.} However there is a standard, skeptical concept of inference or entailment associated with answer set semantics.
This notion of entailment or consequence for programs under the answer set semantics is that a query $Q$ is entailed by a program $\Pi$ if $Q$ is true in all answer sets of $\Pi$, see eg \cite{BGN2000}.
Let us denote this entailment or consequence relation by $\infest_{AS}$. Evidently atoms are true in an answer set if and only if they belong to it.
Conjunctions and disjunctions are handled in the obvious way (eg \cite{LTT99,BGN2000}).
Sometimes, as in \cite{BGN2000}, queries of the form $not\; a$, or in logical notation $\neg a$, are not explicitly dealt with.
However it seems to be in keeping with the semantics to regard a formula of form $\neg a$ to be true in an answer set if and only if $\alpha$ is not true.
Another way to express this would be to say that an answer set satisfies $\neg \alpha$ if it does not violate the constraint $\{ \leftarrow \alpha \}$, where constraint violation is understood as in \cite{LTT99}.%
\footnote{In logical terms this constraint would be written $\alpha \to \bot$.}
In this way we would say that $\Pi \infest_{AS} \neg \alpha$ if no answer set of $\Pi$ contains $\alpha$.
%Similarly, the interpretation of queries containing quantifiers in answer set semantics should also conform to that of equilibrium logic, taking account of any specific restrictions, such as Herbrand models, that might be imposed.

We can therefore transfer interpolation properties from equilibrium logic to answer set semantics and ASP. It remains to consider whether $\infest_{AS}$ is best identified with the closed world version of inference, $\infest_{cw}$, or the more open world version, $\infest_{ow}$. Again, since ASP solvers do not implement inference engines, the difference is really a theoretical one. In traditional logic programming, however, a query that does not belong to the language of the program is usually answered \emph{false}. It also seems quite natural in an ASP context that, given a program $\Pi$ and a query $Q$, one should consider the stable models of $\Pi$ in the language $V (\Pi ) \cup V(Q)$ even if this is a proper extension of the language of $\Pi$.\footnote{Notice that for a non-ground, safe program an atomic query $q(a)$ is automatically false if $a$ does not belong to the language of the program (even if $q$ does), simply because grounding with the program constants is sufficient to generate all answer sets.} So in general $\infest_{cw}$ seems a natural choice for answer set inference. On the other hand, there are contexts where answer set semantics is used in a more open world setting, for example in the setting of hybrid knowledge bases \cite{rosa-05b} where non-monotonic rules are combined with ontologies formalised in description logics. For such systems a semantics in terms of equilibrium logic is provided in \cite{BPPV07}. Here an entailment relation in the style of $\infest_{ow}$ might sometimes be more appropriate.

In general answer set semantics is defined only for coherent programs or theories. For finite, ground, coherent programs, by identifying $\infest_{AS}$ with $\infest_{cw}$, we can apply Proposition \ref{prop:interpol} directly:
\begin{corollary}\label{AS_eq}
For coherent formulas $\alpha$, $(\infest, \infest )$-interpolation in the form of Proposition \ref{prop:interpol} holds for entailment $\infest_{AS}$ in answer set semantics.
\end{corollary}

\section{Uniform Interpolation and Forgetting}

A stronger form of interpolation known as \emph{uniform} interpolation is also important for certain applications in computer science. As usual, given $\alpha, \beta$ with $\alpha \vdash \beta$, we are interested in interpolants $\gamma$ such that
\begin{equation}
\label{u-interpolation}
\alpha  \vdash \gamma  \;\; \& \;\;  \gamma \vdash \beta
\end{equation}
\change{where $V(\gamma) \subseteq V(\alpha) \cap V(\beta)$.}
% where $\gamma$ contains only predicate and constant
% symbols that belong to both $\alpha$ and $\beta$.
The difference now is that $\gamma$ is said to be a \emph{uniform interpolant} if (\ref{u-interpolation}) holds \emph{for any} $\beta$ in the same
%signature 
\change{language} such that $\alpha \vdash \beta$. A logic is said to have the uniform interpolation property if such uniform interpolants exist for all $\alpha, \beta$.

While uniform interpolation fails in classical logic and in many non-classical logics, it may hold when certain restrictions are placed on the theory language where $\alpha$ is formulated and the query language containing $\beta$. For example it has been shown to hold for some description logics (\cite{KWZ08}) where such syntactic restrictions apply. Even in ASP it turns out that a form of uniform interpolation holds for a very restricted query language, essentially one that allows just instance retrieval. We can show this by using some known results in ASP about the concept of \emph{forgetting} \cite{EW08} that is quite closely related to interpolation.

Variable forgetting, as studied in \cite{EW08}, is concerned with the following problem. Given a disjunctive logic program $\Pi$ and a certain atom $a$ occurring in $\Pi$, construct a new program, to be denoted by ${\bf forget}(\Pi, a)$, that does not contain $a$ but whose answer sets are in other respects as close as possible to those of $\Pi$. For the precise notion of closeness the reader is referred to \cite{EW08}, however some consequences will be evident shortly. In \cite{EW08} the authors define ${\bf forget}(\Pi, a)$ (as a generic term), show that such programs exist whenever $\Pi$ is coherent, and provide different algorithms to compute such programs.

Given coherent $\Pi$ and $a$ in $\Pi$, the results ${\bf forget}(\Pi, a)$, of forgetting about $a$ in $\Pi$ may be different but are always answer set equivalent. Moreover for our purposes they satisfy the following key property, where $\Pi$ is coherent, $a,b$ are distinct atoms in $\Pi$ and as usual $\infest$ denotes nonmonotonic consequence,
\begin{equation}
\label{forget}
\Pi \infest b \Leftrightarrow {\bf forget} (\Pi, a) \infest b .
\end{equation}
showing that indeed the answer sets of $\Pi$ and ${\bf forget}(\Pi, a)$ are closely related.

To establish a version of uniform interpolation for the case of disjunctive programs and simple, atomic queries, we need to show that we can always find a $\Pi' = {\bf forget}(\Pi, a)$ such that $\Pi \infest \Pi'$. For this we can examine the first algorithm of \cite{EW08} for computing ${\bf forget}(\Pi, a)$; this is also the simplest of the three algorithms presented. Let $\Pi$ be a coherent program with rules of form (\ref{disj_rule}) that we write as formulas of form (\ref{disj_formula}) and let $a$ be an atom in $\Pi$. The method for constructing a $\Pi' = {\bf forget}(\Pi, a)$ is as follows.
\begin{enumerate}
\item Compute the equilibrium models $E(\Pi )$.
\item Let $E'$ be the result of removing $a$ from each $\mathcal{M} \in E(\Pi)$.
\item Remove from $E'$ any model that is non-minimal to form $E'' (= \{ A_1 , \ldots , A_m \}$, say).
\item Construct a program $\Pi'$ whose answer sets are precisely $\{ A_1 , \ldots , A_m \}$ as follows:
\begin{itemize}
\item for each $A_i$, set $\Pi_i = \{ \neg \overline{A}_i \to a'\colon a' \in A_i \}$, where $\overline{A}_i = V(\Pi) \smallsetminus A_i$.
\item Set $\Pi' = \Pi_1 \cup \ldots \cup \Pi_m$.
\end{itemize}
\end{enumerate}
We can now verify the desired property. Let $\mathcal{L}$ be the simple query language composed of conjunctions of literals.

\begin{proposition}
\label{prop:u-interpolation}
In equilibrium logic (or answer set programming) uniform interpolation holds for (coherent) disjunctive programs and queries in $\mathcal{L} (V(\Pi))$.
\end{proposition}
\begin{zproof}
To prove the claim we shall show the following. Let $\Pi$ be a coherent disjunctive program and let $\mathcal{L}$ = $\mathcal{L}(V)$ for some $V \subseteq V(\Pi)$. Then there is a program $\Pi'$ such that $V(\Pi') = V$ and for any $\varphi \in \mathcal{L}$,
\begin{equation}
\label{interpol}
\Pi \infest \varphi \to (\Pi \infest \Pi' \;\; \& \;\; \Pi' \infest \varphi )
\end{equation}

To begin, let $\Pi$ and $\varphi$ be as above with $\Pi \infest \varphi$. Let $X = \{ a_1 , \ldots , a_n \} = V(\Pi) \smallsetminus V$. Then we
choose $\Pi'$ to be the result of forgetting about $X$ in $\Pi$, defined in \cite{EW08} as follows:
\[
{\bf forget}(\Pi, X) := {\bf forget}({\bf forget}({\bf forget}(\Pi, a_1), a_2), \ldots , a_n ), 
\]
and it is shown there that the order of the atoms in $X$ does not matter. Now we know by (\ref{forget}) that for any atom $a \in V$ and any $i = 1,n$, 
\begin{equation}
\label{forgeta}
\Pi \infest a \Leftrightarrow {\bf forget}(\Pi, a_i) \infest a, 
\end{equation}
therefore 
\begin{equation}
\label{forgetX}
\Pi \infest a \to {\bf forget}(\Pi, X) \infest a. 
\end{equation}
Let $\Pi'$ be ${\bf forget}(\Pi, X)$ as determined by algorithm 1 of \cite{EW08} described above. It is easy to see by (\ref{forgeta}) and the semantics of $\infest$ that  (\ref{forgetX}) continues to hold where $a$ is replaced by a negated atom $\neg b$ and therefore also by any conjunction of literals since a conjunction is entailed only if each element holds in every equilibrium model.\footnote{As \cite{EW08} points out, if an atom $b$ is true in some answer set of ${\bf forget}(\Pi, a)$, then it must also be true in some answer set of $\Pi$, showing that (\ref{forgetX}) holds for literals.} So it remains to show that $\Pi \infest \Pi'$. Again, it will suffice to  show this entailment for one member of the sequence 
${\bf forget}(\Pi, a_i)$ and since the order is irrelevant wlog we can choose the first element ${\bf forget}(\Pi, a_1)$ and show that $\Pi \infest {\bf forget}(\Pi, a_1)$. We compute the programs $\Pi_1 , \ldots , \Pi_m$ as in the algorithm. Then we need to check that $\Pi \infest \Pi_i$ for any $i = 1,\dots,n$, ie that for each $\mathcal{M} \in E(\Pi)$, $\mathcal{M} \models \{ \neg \overline{A}_i \to a' : a' \in A_i \}$. 

Consider $\mathcal{M} \in E(\Pi)$ where $\mathcal{M} = \langle T,T \rangle $. We distinguish two cases. (i) $A_i \subseteq T$. Then $\mathcal{M} \models a'$ for each $a' \in A_i$. It follows that $\mathcal{M} \models \neg \overline{A}_i \to a'$ for each $a' \in A_i$ and so 
$\mathcal{M} \models \{ \neg \overline{A}_i \to a' : a' \in A_i \}$. Case (ii) $A_i \not \subseteq T$. Then $T$ and $A_i$ are incomparable. In particular we cannot have $T \subset A_i$ by the minimality property of $A_i$ obtained in step~3. Hence $T \cap \overline{A}_i \not = \emptyset$. Choose $a'' \in T \cap \overline{A}_i $. Then $\mathcal{M} \models a''$, so $\mathcal{M} \not \models \neg a''$ and hence $\mathcal{M} \not \models \neg \overline{A}_i$. Consequently, for any $a'$,
$\mathcal{M}  \models \neg \overline{A}_i \to a'$ and so 
$\mathcal{M} \models \{ \neg \overline{A}_i \to a' : a' \in A_i \}$.

It follows that for any $i$, $\Pi \infest \Pi_i$ and so by construction $\Pi \infest \Pi'$, which establishes the proposition.
\end{zproof}

\subsection{Extending the query language}

If we establish uniform interpolation in ASP using the method of forgetting, as defined in \cite{EW08}, it seems clear that we cannot extend in a non-trivial way the expressive power of the query language $\mathcal{L}$. Since the method of forgetting $a$ in $\Pi$ removes non-minimal sets from $E(\Pi)$ (once $a$ has been removed), an atom $b$ might be true in some equilibrium model of $\Pi$ but not in any equilibrium model of ${\bf forget}(\Pi, a)$. Hence we might have a disjunction, say $a \lor b$, derivable from $\Pi$ but not from ${\bf forget}(\Pi, a)$.
% Likewise, if we consider programs with variables
% in a first-order setting, we cannot in general 
% extend $\mathcal{L}$ to include existential queries. 

On the other hand, the property of uniform interpolation certainly holds for any $\mathcal{L}(V)$ even without the condition that $V \subseteq V(\Pi)$. Suppose that $\Pi \infest \varphi$ where $V(\varphi) \smallsetminus V(\Pi) \not = \emptyset$, say 
$V(\varphi) \smallsetminus V(\Pi)  = \{ b_1 , \ldots , b_k \}$. Then $ b_1 , \ldots , b_k$ are false in all equilibrium models of $\Pi$. Trivially, if $b$ is not in $V(\Pi)$ we can regard the result of forgetting about $b$ in $\Pi$ as just $\Pi$. So we can repeat the proof of Proposition \ref{prop:u-interpolation}, but now setting $X = \{ V(\Pi) \smallsetminus V \} \cup \{ V \smallsetminus V(\Pi) \}$. All the relevant properties will continue to hold.

An interesting open question is whether we can extend the theory language to include more general kinds of program rules  allowing negation in the head. Accommodating these kinds of formulas would constitute an important generalisation since they amount to a normal form in equilibrium logic. However, the answer sets of such programs do not satisfy the minimality property that holds for the answer sets of disjunctive programs, so it is clear that the definition of forgetting would need to be appropriately modified - a task that we do not attempt here.

\section{Literature and Related Work}
The interpolation theorem for classical logic is due to Craig~\cite{Cr57}; it was extended to intutionistic logic by Sch\"utte~\cite{Sc62}. Maksimova~\cite{Ma77} characterised the super-intuitionistic propositional logics possessing interpolation. A modern, comprehensive treatment of interpolation in modal and intuitionistic logics can be found in the monograph~\cite{GM2005} by Gabbay and Maksimova. 

In non-monotonic logics, interpolation has received little attention. A notable exception is an article~\cite{Am02} by Amir establishing some interpolation properties for circumscription and default logic. By the well-known relation between the answer sets of disjunctive programs and the extensions of corresponding default theories, he also derives a form of interpolation for ASP. With regard to answer set semantics, the approach of \cite{Am02} is quite different from ours. Since it is founded on an analysis of default logic, it uses classical logic as an underlying base. So Amir's version of interpolation is a form of (\ref{eq:interpol1}) where $L$ is classical logic; there is no requirement that $\vdash_L$ form a well-behaved sublogic of $\infest$, eg a deductive base. As Amir remarks, one cannot deduce in general from property (\ref{eq:interpol2}) that $\alpha \infest \beta$. However if $L$ is classical logic one cannot even deduce $\alpha \infest \beta$ from (\ref{eq:interpol1}). More generally, there is no counterpart to our Proposition \ref{prop:interprops} in this case. Another difference with respect to our approach is that \cite{Am02} does not discuss the nature of the $\infest$ relation for ASP in detail, in particular how to understand $\Pi \infest \varphi$ in case $\varphi$ contains atoms not present in the program $\Pi$. In fact, if we interpret $\infest_{AS}$ as in section 5 above, it is easy to refute $(\infest , \vdash_L)$-interpolation where $L$ is classical logic. Let $\Pi$ be the program $B \leftarrow \neg A$ and $q$ the query $B \land \neg C$. Then clearly $\Pi \infest_{AS} q$, but there is no formula in the vocabulary $B$ that would classically entail $\neg C$. Under any interpretation of answer set inference such that atoms not in the program are regarded as false, $(\infest , \vdash_L)$-interpolation would be refuted.

\section{Conclusions}
We have discussed two kinds of interpolation properties for non-monotonic inference relations and shown that these properties hold in turn for the two different inference relations that we can associate with propositional equilibrium logic. In each case we use the fact that the collection of equilibrium models is definable in the logic $\mathbf{HT}$ of here-and-there and that this logic possesses the usual form of interpolation. One of the forms of inference studied seems to be in many cases an appropriate concept to associate with answer set programming, although in general ASP systems are not tailored to query answering or deduction. Using results from \cite{EW08} about variable forgetting in ASP, we could also show how the property of uniform interpolation holds for disjunctive programs and a restricted query language.

A forthcoming, extended version of this paper will deal with interpolation in quantified equilibrium logic and non-ground ASP. Results similar to the propositional case can be established, providing that the class of equilibrium models is (first-order) definable, as for instance in the case of safe theories.

\bibliographystyle{plain}

\end{document}